\documentclass[prd,preprint,nofootinbib,longbibliography]{revtex4-1} 

\usepackage{latexsym,verbatim}
\usepackage{amssymb}
\usepackage{amsfonts}
\usepackage{amsmath,color}
\usepackage[draft]{graphicx}
\usepackage{bm}
\usepackage[utf8]{inputenc}
\usepackage[T1]{fontenc}

\usepackage{hyperref}

\hypersetup{
    colorlinks=true,
    linkcolor=red,
    citecolor=red,
    filecolor=magenta,      
    urlcolor=cyan,
    pdftitle={GEM Review},
    pdfpagemode=FullScreen,}

\def\mb#1{\mathbf{#1}}

\def\ber{\begin{eqnarray}}
\def\eer{\end{eqnarray}}
\def\beq{\begin{equation}}
\def\eeq{\end{equation}}

\def\rmd{{\rm d}}

\def\ed{\end{document}}

\def\bfbeta{\pmb{\beta}}

\makeatletter
\let\jnl@style=\rm
\def\ref@jnl#1{{\jnl@style#1}}

\def\aj{\ref@jnl{AJ}}                   
\def\actaa{\ref@jnl{Acta Astron.}}      
\def\araa{\ref@jnl{ARA\&A}}             
\def\apj{\ref@jnl{ApJ}}                 
\def\apjl{\ref@jnl{ApJ}}                
\def\apjs{\ref@jnl{ApJS}}               
\def\ao{\ref@jnl{Appl.~Opt.}}           
\def\apss{\ref@jnl{Ap\&SS}}             
\def\aap{\ref@jnl{A\&A}}                
\def\aapr{\ref@jnl{A\&A~Rev.}}          
\def\aaps{\ref@jnl{A\&AS}}              
\def\azh{\ref@jnl{AZh}}                 
\def\baas{\ref@jnl{BAAS}}               
\def\bac{\ref@jnl{Bull. astr. Inst. Czechosl.}}
\def\caa{\ref@jnl{Chinese Astron. Astrophys.}}
\def\cjaa{\ref@jnl{Chinese J. Astron. Astrophys.}}
\def\icarus{\ref@jnl{Icarus}}           
\def\jcap{\ref@jnl{J. Cosmology Astropart. Phys.}}
\def\jrasc{\ref@jnl{JRASC}}             
\def\memras{\ref@jnl{MmRAS}}            
\def\mnras{\ref@jnl{MNRAS}}             
\def\na{\ref@jnl{New A}}                
\def\nar{\ref@jnl{New A Rev.}}          
\def\pra{\ref@jnl{Phys.~Rev.~A}}        
\def\prb{\ref@jnl{Phys.~Rev.~B}}        
\def\prc{\ref@jnl{Phys.~Rev.~C}}        
\def\prd{\ref@jnl{Phys.~Rev.~D}}        
\def\pre{\ref@jnl{Phys.~Rev.~E}}        
\def\prl{\ref@jnl{Phys.~Rev.~Lett.}}    
\def\pasa{\ref@jnl{PASA}}               
\def\pasp{\ref@jnl{PASP}}               
\def\pasj{\ref@jnl{PASJ}}               
\def\rmxaa{\ref@jnl{Rev. Mexicana Astron. Astrofis.}}%
\def\qjras{\ref@jnl{QJRAS}}             
\def\skytel{\ref@jnl{S\&T}}             
\def\solphys{\ref@jnl{Sol.~Phys.}}      
\def\sovast{\ref@jnl{Soviet~Ast.}}      
\def\ssr{\ref@jnl{Space~Sci.~Rev.}}     
\def\zap{\ref@jnl{ZAp}}                 
\def\nat{\ref@jnl{Nature}}              
\def\iaucirc{\ref@jnl{IAU~Circ.}}       
\def\aplett{\ref@jnl{Astrophys.~Lett.}} 
\def\apspr{\ref@jnl{Astrophys.~Space~Phys.~Res.}}
\def\bain{\ref@jnl{Bull.~Astron.~Inst.~Netherlands}}
\def\fcp{\ref@jnl{Fund.~Cosmic~Phys.}}  
\def\gca{\ref@jnl{Geochim.~Cosmochim.~Acta}}   
\def\grl{\ref@jnl{Geophys.~Res.~Lett.}} 
\def\jcp{\ref@jnl{J.~Chem.~Phys.}}      
\def\jgr{\ref@jnl{J.~Geophys.~Res.}}    
\def\jqsrt{\ref@jnl{J.~Quant.~Spec.~Radiat.~Transf.}}
\def\memsai{\ref@jnl{Mem.~Soc.~Astron.~Italiana}}
\def\nphysa{\ref@jnl{Nucl.~Phys.~A}}   
\def\physrep{\ref@jnl{Phys.~Rep.}}   
\def\physscr{\ref@jnl{Phys.~Scr}}   
\def\planss{\ref@jnl{Planet.~Space~Sci.}}   
\def\procspie{\ref@jnl{Proc.~SPIE}}   

\makeatother

\begin{document}

\author{Matteo Luca Ruggiero}
\email{matteoluca.ruggiero@unito.it}
\affiliation{Dipartimento di Matematica ``G.Peano'', Universit\`a degli studi di Torino, Via Carlo Alberto 10, 10123 Torino, Italy}
\affiliation{INFN - LNL , Viale dell'Universit\`a 2, 35020 Legnaro (PD), Italy}
\date{\today}

\title{Synchronization and Fundamental Time: a Connection between Relativity and Quantum Mechanics}

\begin{abstract}
An interesting connection  between special relativity and quantum mechanics was put forward by Louis de Broglie, about 60 years ago, who focused on the link between synchronization in a rotating frame and the quantization of the angular momentum.  Here we generalise his approach to curved spacetime, using the gravitoelectromagnetic {analogy}, which can be applied to describe the weak gravitational field around rotating sources, and give a new interpretation of the results.  \\ 
\end{abstract}

\maketitle

\section{Introduction}\label{sec:intro}

Reconciling quantum mechanics and gravity is one of hardest tasks in theoretical physics. On the one hand, we have  experimental evidences -  culminated in the detection of the Higgs boson \cite{Aad:2012tfa} - supporting the predictions of the Standard Model of particle physics, according to which the ultimate nature of matter is  made of \textit{discrete} entities. On the other hand, the successes of Einstein's theory \cite{Will:2014kxa} testify that General Relativity is our best model to understand gravity, which appears to be the geometrical structure underlying the spacetime \textit{continuum}.  It seems that these two paradigms cannot be reconciled without postulating a discreteness for spacetime too.  It is useful to remember that Einstein himself was well aware of this dichotomy: ``The problem seems to me how one can formulate statements about a discontinuum without calling upon a continuum (spacetime) as an aid; the latter should be banned from the theory as a supplementary construction not justified by the essence of the problem, which corresponds to nothing real. But we still lack the mathematical structure unfortunately'' (quoted by \citet{stachel1986einstein}).

Beyond these fundamental reasons, there are several arguments to support spacetime discreteness (see e.g.  \citet{henson2008causal} and references therein): for instance, the infinities arising when we try  to treat gravity as a quantum field, or the requirement of the finiteness of black hole entropy. Accordingly, various models were considered on the road to a quantum theory of gravity.  For example, in loop quantum gravity \cite{rovelli2008loop} spacetime is discretized at Planck scale and the evolution of spin networks {develops} in discrete steps. In the causal set approach \cite{bombelli1987space},  spacetime is locally described as a finite set of elements with a causal order that enables to define causal connections. 

There were several instances {supporting} a discrete approach: during the 80's, T.D. Lee \cite{lee1983can} suggested that difference equations are more fundamental than differential equations; in his approach space and time are  discretized dynamical variables. 
{Indeed, the proposal to replace differential equations by difference equations in quantum mechanics was already put forward many years before by \citet{diff1}; in addition, a theory of discrete spacetime was worked out by \citet{diff2}, with applications to the electromagnetic field \cite{diff3}.}
A   strategy based on  a finite difference theory was proposed by Caldirola (see e.g. \citet{farias1997introduction} and references therein) to solve the classical problem of the motion of a charged particle in an external electromagnetic field occurring in Abraham-Lorentz and Dirac approaches: in particular, a quantum of time - the so-called \textit{chronon} - was introduced (the numerical value of the chronon is of the same order as the time spent by light to travel an electron classical diameter).

In this context, it is interesting  to point out that in 1959 Louis de Broglie \cite{debroglie1959deux} wrote a note where he focused on the connection between relativity, quantum mechanics:  he showed that the angular momentum quantization is a consequence of the discontinuity of time in a rotating frame (the so-called \textit{time gap} or \textit{synchronization gap} \cite{rizziserafini}) once that a fundamental time scale is assumed. Similar conclusions were achieved, more recently, by \citet{bock2007radial}.  The purposes of this paper is to rephrase de Broglie's argument in order to extend it to curved spacetime. To this end, we are going to use the gravitoelectromagnetic approach: we refer to the fact that  Einstein equations in the weak-field approximation  can be written in analogy with Maxwell equations for the electromagnetic field, where the mass density and current play the role of the charge density and current, respectively \cite{Ruggiero:2002hz,Mashhoon:2003ax,Ruggiero:2023ker}. As an example, we will apply de Broglie's argument to the gravitational field of a slowly rotating source; subsequently, we will discuss a formal generalization.

The paper is organized as follows: in Section \ref{sec:flat} we discuss de Broglie's approach for a rotating frame in flat spacetime; then we generalise it to curved spacetime in Section \ref{sec:gm}. Conclusions are eventually in Section \ref{sec:conc}.

\section{Synchronization gap and angular momentum quantization in flat spacetime}\label{sec:flat}

Rotating frames in relativity have a well known feature: a global Einsteinian clock synchronization is not possible. This fact has an important experimental consequence: the Sagnac effect. Roughly speaking, the latter is a consequence of the  asymmetry in the propagation times of two signals circulating in opposite directions in a rotating reference frame \cite{RR2004,ruggiero2014note,ruggiero2015note}. In order to explain desynchronization, we consider the
line-element in the form\footnote{We use the following notation: Greek (running from 0 to 3) and Latin (running from 1 to 3) indices
denote spacetime and spatial components, respectively;
letters in boldface like $\mb x$ indicate spatial vectors.}
\beq
ds^{2}=g_{00}c^{2}dt^{2}+2g_{0i}cdtdx^{i}+g_{ij}dx^{i}dx^{j}, \label{eq:metricastazionaria}
\eeq
{where we use the local spacetime coordinates defined as $x^\mu =\left[ct, x^i\right]$}, and  $g_{\mu\nu}=g_{\mu\nu}(x^{i})$, hence the spacetime metric does not depend on time. The above metric is quite general in its form and, in particular,  it said to be  \textit{non time-orthogonal}, because $g_{0i} \neq 0$: these off-diagonal terms are generally related to the rotational features of the reference frame and to the rotation of the sources of the gravitational field. It is a well known fact that in such a spacetime, events at spatial locations $x^{i}$ and $x^{i}+dx^{i}$ are simultaneous if the difference of time coordinate is $\displaystyle d x^{0}=-\frac{g_{0i}dx^{i}}{g_{00}}$ (see e.g. \citet{landau2013classical}). This is indeed what happens in a rotating frame: the  line element in a uniformly rotating frame of reference in flat Minkowski spacetime is written (using cylindrical coordinates $\left[ct,r,\varphi,z\right]$ adapted to  the frame) in the form \cite{RR2004}
\beq
ds^{2}=-\left(1-\frac{\omega^{2}r^{2}}{c^{2}} \right)c^{2}dt^{2}+2\frac{\omega r^{2}}{c}d\varphi cdt+dr^{2}+r^{2}d\varphi^{2}+dz^{2}, \label{eq:rot0}
\eeq
where $\omega$ is the constant rotation rate. 

In order to better understand what happens when we try to synchronize clocks in the rotating frame, we consider two different procedures \cite{ashby2003relativity}.  To begin with, we show how  Einstein's synchronization  - which is based on the transmission of light signals - can be used to synchronize clocks in the spacetime described by the line element (\ref{eq:rot0}). Since light signals propagate along null world-lines, we set $ds^{2}=0$ and, for the sake of simplicity, we consider only linear terms in the small parameter $\displaystyle \frac{\omega r}{c}$. Accordingly, from (\ref{eq:rot0}) we get
\beq
dt= \frac{dl}{c}+\frac{\omega r^{2}d\varphi}{c^{2}}, \label{eq:dt1}
\eeq
where we set $dl^{2}=dr^{2}+r^{2}d\varphi^{2}+dz^{2}$. We see that the observers in the rotating frame should add the extra term $\displaystyle \frac{\omega r^{2}d\varphi}{c^{2}}$ to synchronize clocks that are separated by the infinitesimal coordinate distance $dl$.

The same result can be obtained using a quite different synchronization procedure,  based on the use of a portable clock which is slowly transported to synchronize clocks at different spatial locations. As far as the clocks moves, its proper time increment is given by
\beq
d\tau^{2}=-\frac{ds^{2}}{c^{2}}=dt^{2}\left(1-\frac{\omega^{2}r^{2}}{c^{2}}-\frac{2\omega r^{2}}{c^{2}}\frac{d\varphi}{dt}-\frac{1}{c^{2}}\frac{dl^{2}}{dt^{2}} \right). \label{eq:dt2}
\eeq

By hypothesis, the clock is \textit{slowly} moving, hence we may neglect the term $\displaystyle \frac{1}{c^{2}}\frac{dl^{2}}{dt^{2}}$ in Eq. (\ref{eq:dt2}). As a consequence, up to linear order we obtain
\beq
dt= d\tau+\frac{\omega r^{2} d\varphi}{c^{2}}.  \label{eq:dt21}
\eeq
We see again that the additional contribution $\displaystyle \frac{\omega r^{2}d\varphi}{c^{2}}$ should be added to the proper time interval elapsed during the clock motion. 

Eqs. (\ref{eq:dt1}) and (\ref{eq:dt21}) show that in both cases,  path-depending terms (deriving from the rotation of the frame) must be added to what one would  consider in an inertial frame.

In particular, if we want to synchronize clocks {along}  a closed line $C$,  a time discontinuity arises (\textit{time gap}) which is strictly related to the Sagnac effect \cite{bergia1998time,RR2004}:  the coordinate time gap is given by
\beq
\Delta t = -\frac 1 c \oint_{C} \frac{g_{0i}dx^{i}}{g_{00}}. \label{eq:formulafond}
\eeq 
It is possible to calculate the gap in terms of {the} proper  time $\tau$ measured by a clock at rest in the rotating frame. Since $d\tau=\sqrt{g_{00}}dt$, if $C$ is a circumference of radius $r$, using the expression (\ref{eq:rot0}) we obtain:
\beq
\Delta \tau=\frac{\omega r^{2}}{c^{2}} \frac{2\pi}{\sqrt{1-\frac{\omega^{2}r^{2}}{c^{2}}}}. \label{eq:tau2}
\eeq
Notice  that the time gap (\ref{eq:tau2}) is half the Sagnac effect \cite{RR2004}.

Eq. (\ref{eq:tau2}) is the starting point of de Broglie's thought experiment, which we are going to briefly summarize and then rephrase. De Broglie suggested to consider a  train uniformly rotating
 with respect to an inertial frame: its wagons are at rest in the rotating frame and make a circle. In the wagons there are identical clocks, with proper  period $T_{0}$:  to synchronize all clocks without ambiguity in reading their  hands, the time gap should be an integer multiple of $T_{0}$.
 
Without losing the physical meaning of de Broglie's argument, we may rephrase it as follows. We suppose that in the rotating frame we have a set of identical physical oscillators, with proper time $T_{0}$ and rest mass $m$; we also suppose that they are placed along a circle of radius $r$. These oscillators define the \textit{fundamental time scale} $T_{0}$. As a consequence, the time gap (\ref{eq:tau2}) should satisfy the following relation: 
\beq
\Delta \tau=\int_{C} \frac{\omega r^{}}{c^{2}} \frac{rd\varphi}{\sqrt{1-\frac{\omega^{2}r^{2}}{c^{2}}}}=nT_{0}, \label{eq:tau3}
\eeq
where $n \in \mathbb{N}$. In other words, the time gap is a an integer multiple of the fundamental time scale. {de Broglie   hypothesizes that each oscillator  can be thought of as a particle of rest mass $m$; then according to the law of quantum mechanics, a proper frequency  $\nu=\frac{1}{T_{0}}$ can be associated to it,  in such a way that $mc^{2}=h\nu$. We remark that this should be considered as a thought experiment, rather than an operational definition of a fundamental time scale.}  Hence, from (\ref{eq:tau3}) we obtain
\beq
\int_{C}  \frac{m \omega r^{2} d\varphi}{\sqrt{1-\frac{\omega^{2}r^{2}}{c^{2}}}}=n h. \label{eq:tau4}
\eeq
But  $\displaystyle \ell= \frac{m\omega r^{2}}{\sqrt{1-\frac{\omega^{2}r^{2}}{c^{2}}}}$ is the angular momentum of each oscillator with respect to the centre of circle, so we may write
\beq
\int_{C}  \ell d\varphi=n h, \label{eq:quantum1}
\eeq
which is the quantization {rule of  the angular momentum}. Eq. (\ref{eq:quantum1}) can be written  using the kinetic momentum $p_{\varphi}=\frac{m \omega r}{\sqrt{1-\frac{\omega^{2}r^{2}}{c^{2}}}}$ in the form
\beq
\int_{C}  p_{\varphi } r d\varphi=n h. \label{eq:quantum2}
\eeq
According to de Broglie, this result suggests a deep connection between the (non unique) definition of time in relativity and quantum mechanics. 
\section{Generalization to curved spacetime}\label{sec:gm}

In order to extend the previous results to a general relativistic framework, we focus on the gravitational field of a rotating isolated source. 
 In the weak-field and slow-motion approximation, it is possible to describe this field using the gravitoelectromagnetic  approach: in fact, Einstein's equations in this approximation  can be written in analogy with electromagnetism (see e.g.  \citet{,Ruggiero:2002hz,Mashhoon:2003ax,Ruggiero:2023ker}). Accordingly, the corresponding spacetime is described by the following line element
 \begin{equation} 
ds^2=-c^2\left(1-2\frac{\Phi}{c^2}\right)dt^2-\frac{4}{c}({\bf 
A}\cdot d{\bf
x})dt+\left(1+2\frac{\Phi}{c^2}\right) \delta_{ij}dx^idx^j, \label{eq:metrica1}
\end{equation}
where the gravitoelectric potential $\Phi$ and the gravitomagnetic potential $\mb A$ are related to the source through its mass $M$ and angular momentum $\mb J$
\begin{equation}\label{eq:MJ} \Phi = \frac{GM}{r},\quad {\bf A}=
\frac{G}{c}\frac{{\bf J}\wedge {\bf
x}}{r^3}. 
\end{equation}
In the above equations $\bf x$ is the position vector and $r=|\bf x|$. 

The above metric (\ref{eq:metrica1})  is in the general form (\ref{eq:metricastazionaria}): in particular, we have ${g}_{0i}=-2A_i/c^2$.  It can be written in the axially symmetrical  form using cylindrical coordinates $\left[ct,r,\varphi,z\right]$  in the equatorial plane of the source (where $z=0$ and we suppose that $\mb J$ is perpendicular to the equatorial plane; in addition we remember that  $g_{\varphi\varphi}=r^{2}$):
\beq
ds^{2}=g_{00}c^{2}dt^{2}+2g_{0\varphi} d\varphi c dt+g_{\varphi\varphi}d\varphi^{2}. \label{eq:metrica2}
\eeq
These coordinates are adapted to an inertial frame at rest with respect to the centre of mass of the source. We consider a frame uniformly rotating with respect to the inertial frame, so that $\varphi=\phi+\omega t$, where $\omega$ is the (constant) rotation rate; consequently, from (\ref{eq:metrica2}) we obtain
\beq
ds^{2}=c^{2}dt^{2}G_{00}+2G_{0\phi}d\phi c dt+r^{2}d\phi^{2},
\eeq
where $\displaystyle G_{00}=-\left(1-\frac{2\Phi}{c^{2}}-\frac{\omega^{2}r^{2}}{c^{2}}-\frac{2g_{0\varphi }\omega}{c} \right)$ and $ \displaystyle G_{0\phi}=\frac{\omega r^{2}}{c}+g_{0\varphi}$.

In this frame, it possible to evaluate the  coordinate time  gap (\ref{eq:formulafond}):
\beq
\Delta t =- \frac 1 c \int_{C}  \frac{G_{0\phi}}{G_{00}} d\phi. \label{eq:tau5}
\eeq
{Using} the above expressions for $G_{00},G_{0\phi}$, and setting $\gamma=\frac{1}{\sqrt{|G_{00}|}}$, we obtain  the proper time gap for an observer at rest in the rotating frame:
\beq
\Delta \tau= \int_{C} \frac{\gamma}{c} \left(\frac{\omega r^{2}}{c}+g_{0\varphi} \right)d\phi. \label{eq:tau6}
\eeq

If we proceed as in the case of the rotating disk (the  identical oscillators of mass $m$ and proper time $T_{0}$ are rotating around the source of the gravitational field along a circle of radius $r$ in the plane $z=0$), we obtain
\beq
\int_{C} \gamma \left(m \omega r^{2}+mc g_{0\varphi} \right) d\phi = n h.  \label{eq:tau7}
\eeq
It is $g_{0i}=-2 A_{i}/c^{2}$, hence we get

\beq
\int_{C} \gamma \left(m \omega r^{2}-\frac{2m}{c}A_{\varphi} \right) d\phi = n h.  \label{eq:tau8}
\eeq
Introducing the physical {components} of the gravitomagnetic potential $\hat A_{\varphi}=A_{\varphi}/r$, we may write the above equation in the form\footnote{Up to linear order in the gravitoelectromagnetic potentials.}

\beq
\int_{C}  \left(p_{\varphi}-\frac{2m}{c}\hat A_{\varphi} \right)r  d\phi = n h.  \label{eq:tau9}
\eeq
Eq. (\ref{eq:tau9}) extends the quantization rule (\ref{eq:quantum2})  to curved spacetime, in terms of  the generalized momentum $P_{\varphi}=\left(p_{\varphi}-\frac{2m}{c}\hat A_{\varphi} \right)$:  this expression is in agreement with the gravitoelectromagnetic formalism, since the {gravitomagnetic} charge is $q_{B}=-2m$ (see \citet{Mashhoon:2003ax}). If the oscillators {are} at rest in the inertial frame of the source, we obtain
 \beq
 \int_{C}  \left(\frac{2m}{c}\hat A_{\varphi} \right)r  d\phi = n h.  \label{eq:tau10}
 \eeq

We can give a different interpretation of the above results if we use the gravitoelectromagnetic formalism in its generality. In fact, if we start (in both flat and curved spacetime) from a stationary metric in the form (\ref{eq:metricastazionaria}), we can formally set
\[
\frac{\Phi}{c^{2}}=\frac{g_{00}+1}{2} \quad \frac{\Psi}{c^{2}}=\frac{g_{ij}-1}{2} \quad \frac{A_{i}}{c^{2}}=-\frac{g_{0i}}{2},
\]
where $|\frac{\Phi}{c^{2}}| \ll 1$, $|\frac{\Psi}{c^{2}}| \ll 1$, $|\frac{A_{i}}{c^{2}}| \ll 1$; accordingly, the metric (\ref{eq:metricastazionaria})  can be written in the form
\beq
\mathrm{d} s^2= -c^2 \left(1-2\frac{\Phi}{c^2}\right)\rmd t^2 -\frac4c A_{i}\rmd x^{i}\rmd t  +
 \left(1+2\frac{\Psi}{c^2}\right)\delta_{ij}\rmd x^i \rmd x^j\ , \label{eq:weakfieldmetric11}
\eeq
It is easy to show (see e.g. \citet{Ruggiero:2021uag}) that if we define  the gravitoelectromagnetic fields as
\beq
\mb B= \bm \nabla \wedge \mb A, \quad \mb E= -\bm \nabla \Phi, \label{eq:defEtime}
\eeq
the geodesic equation to linear order in $\bfbeta={\mathbf v}/c$  takes the form 
 {\beq
\frac{\rmd \mb v}{\rmd t}=-\mb E-2({\pmb \beta}\wedge {\mathbf B}) \label{eq:lor001}
\eeq}
Differently speaking, the gravitoelectromagnetic analogy allows to describe the motion of test particles in terms of the action of a Lorentz-like force equation. Actually, as discussed by \citet{Ruggiero:2023ker}, a similar analogy can be obtained in full theory, using a splitting approach.

So, we see that  Eq. (\ref{eq:tau10}) can be obtained with a certain generality when we start from a metric in the form (\ref{eq:metricastazionaria}): in this case, the oscillators defining the fundamental time scale are at rest with respect to the coordinates used to express the metric.  In particular, if we apply the Stokes theorem  to the lhs of Eq. (\ref{eq:tau10})  we get
\beq
\frac{2m}{c} \int_{C}  \mb A \cdot d \mb x  = \frac{2m}{c} \int_{S} \bm \nabla \wedge \mb A \cdot d\mb S = \frac{2m}{c} \int_{S} \mb B \cdot d\mb S=nh \label{eq:quant}
\eeq
where $\mb S$ is the area vector enclosed by the contour line $C$. In conclusion, the hypothesis of the existence of a fundamental time scale can be interpreted as a quantization of the gravitomagnetic \textit{flux}. Notice that the quanta of the gravitomagnetic flux depend on the mass $m$ of the oscillators defining the fundamental time scale.

\section{Conclusions}\label{sec:conc}

There is an interesting connection between relativity and quantum mechanics which arises when we consider {rotation} effects in relativity. This connection was emphasized  in a short note written in 1959 by de Broglie, who showed that if we assume that the synchronization gap in a rotating reference frame is an integer multiple of a fundamental time scale,  then the quantization rule of the angular momentum is obtained. In other words, if time in a rotating frame is measured by a collection of physical oscillators which define a fundamental time scale, then their angular momentum is quantized. We have shown here that this approach can be generalized to {curved} spacetime and, in particular, 
we have focused on the case of the field around a rotating source. This field, in the weak-field and slow-motion approximation, can be described using the gravitoelectromagnetic analogy: we have shown that in this case the hypothesis of a fundamental time scale implies a quantization rule for the generalized momentum or,  said differently, for the flux of the gravitomagnetic field.  

We believe that,  as de Broglie suggested, these remarks, if further analyzed, could be useful to shed new light on the {relationship} between relativity and quantum mechanics at a fundamental level.

\begin{acknowledgments}
The author acknowledges the contribution of  the local research  project \textit{Modelli gravitazionali per lo studio dell'universo (2022)} - Dipartimento di Matematica ``G.Peano'', Universit\`a degli Studi di Torino; this  work is done within the activity of the Gruppo Nazionale per la Fisica Matematica (GNFM). The author thanks Dr. Antonello Ortolan for  useful discussion.
\end{acknowledgments}


%

\end{document}